\documentstyle[eqsecnum,floats,preprint,epsf,aps]{revtex}

\tighten

\begin{document}

\newcommand{\vp}{\varphi}
\newcommand{\CP}{${\cal CP}$}
\newcommand{\C}{${\cal C}$}
\newcommand{\Parity}{${\cal P}$}
\newcommand{\B}{${\cal B}$}
\newcommand{\be}{\begin{equation}}
\newcommand{\ee}{\end{equation}}
\newcommand{\bea}{\begin{eqnarray}}
\newcommand{\eea}{\end{eqnarray}}
\newcommand{\PSbox}[3]{\mbox{\rule{0in}{#3}\includegraphics{#1}\hspace{#2}}}

\title{Semi-Analytical Approaches to Local Electroweak Baryogenesis}
\author{Arthur Lue$^{1}$\footnote[1]{ithron@mit.edu}, 
Krishna Rajagopal$^{2}$\footnote[2]{krishna@theory.caltech.edu} and 
Mark Trodden$^{1}$\footnote[3]{trodden@ctpa04.mit.edu.~~ Also, Visiting 
Scientist, Brown University, Providence, RI 02912.}}
\address{~\\$^1$Center for Theoretical Physics \\ 
Laboratory for Nuclear Science and Department of Physics \\
Massachusetts Institute of Technology, Cambridge, Massachusetts 02139, USA.}
\address{~\\$^2$Lauritsen Laboratory of High Energy Physics\\
California Institute of Technology, Pasadena, CA 91125, USA.}
\maketitle
\begin{abstract}
We examine two semi-analytical 
methods for estimating the baryon asymmetry of the 
universe (BAU) generated in scenarios of ``local'' electroweak baryogenesis
(in which the requisite baryon number violation 
and \CP\ violation occur together in space and time). 
We work with the standard electroweak theory augmented by the addition of
a \CP\ violating dimension six operator.
We work in the context of a first order phase transition,
but the processes we describe can also occur during the
evolution of a network of topological defects.
Both the approaches we explore deal with circumstances where
the bubble walls which convert the high
temperature phase to the low temperature phase are thin and rapidly moving. 
We first consider the dynamics of
localized configurations with winding number one which remain
in the broken phase immediately after the bubble wall has passed.
Their subsequent decay 
can anomalously produce 
fermions. 
In a prelude to our analysis of this effect, we
demonstrate how to define the
\C\ and \CP\ symmetries in the bosonic sector of the electroweak theory 
when configurations with nonzero winding are taken into account.
Second, we consider the effect 
of the passage of the wall itself on configurations which
happen to be near the crest of the ridge between vacua 
as the wall arrives.
We find that neither
of the simple approaches followed here can be pushed far enough to obtain
a convincing estimate of the BAU which is produced. A large scale
numerical treatment seems necessary.
\end{abstract}
\setcounter{page}{0}
\thispagestyle{empty}
\vfill
\baselineskip 14pt

\noindent MIT-CTP-2590 

\noindent CALT-68-2089

\eject

\vfill

\eject

\baselineskip 24pt plus 2pt minus 2pt

\section{Introduction}

One of the central challenges of modern cosmology is to explain how the
baryon asymmetry of the universe (BAU) can be dynamically generated from 
baryon number symmetric initial conditions. The relevant observational 
quantity is the baryon to entropy ratio which is tightly
constrained by primordial nucleosynthesis to lie in the interval
$2\times 10^{-10} < n_B/s < 5\times 10^{-10}$.
Sakharov\cite{Sakharov} showed that there are three necessary 
conditions which must be satisfied if a particle physics model is
to produce a  net cosmological
baryon asymmetry.
These are a violation of baryon number (\B) conservation, 
a departure from thermal equilibrium, and
violations of charge conjugation (\C) and charge-parity (\CP) 
symmetries. 
The standard model of electroweak interactions includes 
all three ingredients, as we discuss below.
This observation has 
led to the suggestion that the standard model might be responsible for the 
generation of the baryon asymmetry of the 
universe
\cite{DS,KRS,CKN1,CKN2,TZ,DHSS,CKN3,CKN4,F&S,GS,JPT,CKV,HN,DMEWBG,review}.

Baryon number is exactly conserved at
the classical level in the standard electroweak theory,
but at the quantum level this ceases to be true as a 
consequence of the anomaly\cite{tHooft}.
The relevant process at zero temperature is an
exponentially suppressed tunnelling under the 
energy barrier which separates
inequivalent vacua in gauge 
and Higgs field 
configuration space. 
The lowest point on the crest of the barrier is a 
saddle point configuration called the sphaleron\cite{sphaleron}.
At nonzero temperature and in particular for
temperatures around or above the critical temperature of the electroweak 
phase transition it is possible for thermal effects to cause classical 
transitions over the sphaleron barrier\cite{KRS,AM}. Although at 
temperatures above the 
electroweak phase transition anomalous baryon number 
and lepton number (${\cal L}$) violation is
unsuppressed in the standard model, this occurs in such a way that 
${\cal B}-{\cal L}$ is conserved. Thus, any preexisting ${\cal B}-{\cal L}$
asymmetry cannot be erased by electroweak processes.
We are assuming that above the temperature of the 
electroweak phase transition ${\cal B}$ and ${\cal L}$ are both zero.

In order to construct a scenario for electroweak baryogenesis we must achieve 
a departure from thermal equilibrium. In the context of the early
universe this is typically satisfied in one of two ways. One possibility is
that the 
electroweak phase transition is considered to be first order so that 
the violent conversion of the high temperature phase to 
the low temperature phase results in
non-equilibrium conditions near the bubble 
walls separating the phases\cite{CKN2,TZ,DHSS,CKN3,CKN4,F&S,GS,JPT,CKV,HN}. 
An alternative is that non-equilibrium processes
occur in topological 
defects, such as cosmic strings, around which the electroweak symmetry is 
restored\cite{DMEWBG}. In defect scenarios, the phase boundaries between the 
interior and exterior of the defects behave in an analogous manner to bubble 
walls. 
In this paper, we shall phrase our considerations entirely in terms of bubble
walls at a first order phase transition, even though the 
processes we discuss also arise in defect mediated baryogenesis.

We make several standard assumptions related to the 
rate per unit volume $\Gamma$ for 
baryon number violating processes in thermal equilibrium,
which we write as
\begin{equation}
\Gamma = \kappa (\alpha_W T)^4 \ .
\label{kappa}
\end{equation}
We assume that the phase transition is strongly enough first
order that at temperatures below $T_c$, 
$\Gamma$ is small enough 
that
the BAU created at the transition  is not 
erased by subsequent electroweak baryon number violating processes.
(The constraint turns out to be $\kappa < 10^{-9}$
for $T<T_c$\cite{nowashout,KLRS}.)
It now seems likely that this requires 
extending the field content of the theory
in some way which makes the phase transition more
strongly first order.  As is often done, we use the minimal standard model
with a Higgs mass lighter than is allowed by experiment
to model the strongly first order phase transition in the
extended theory.
We also assume, again as is standard, that at temperatures
just above $T_c$ the dimensionless quantity $\kappa$ is not very small.

Sakharov's third condition is
that \C\ and \CP\ must be violated. The standard electroweak
theory is maximally \C\ violating due to the V-A nature of the interactions.
As we will show, the purely bosonic sector of the theory is also \C\ violating
once topological properties of the sector are treated correctly. 
The only violation of \CP\ in the electroweak sector of the standard model 
occurs in the CKM matrix.
Electroweak baryogenesis scenarios using this source of \CP\ violation are
generally expected to produce a BAU which is far too small.
We can also ignore the \CP\ 
violating $\theta$ term of QCD since this is known to be tiny.
Therefore, if we are to generate an appreciable baryon
asymmetry 
it is necessary to consider extending the standard model to include new
sources of \CP\ violation.
A popular approach has
been to consider a two-Higgs doublet model with  
explicit renormalizable \CP\ violating terms
in the Higgs 
sector\cite{TZ,CKN3,CKN4,JPT,CKV,DMEWBG}. This
enhances the \CP\ violation beyond that present in the standard
model
but adds a host of new free parameters to the theory.
Here we follow the simpler
approach first discussed by Dine {\it et al.}\cite{DHSS}.
We assume that there is some \CP\ violating
physics (beyond that in the standard model) at an
energy scale $M$ greater than $v=250~{\rm GeV}$. At scales
less than $M$, the effective theory is the standard model
plus nonrenormalizable operators, some of
which are \CP\ odd.  In this paper, for definiteness,  we add the 
dimension six \CP\ odd operator
\be
{\cal O}=\frac{b}{M^2}\,\hbox{Tr}(\Phi^{\dagger}\Phi)
\hbox{Tr}(F_{\mu\nu} \tilde{F}^{\mu\nu}) \ ,
\label{newop}
\ee
to the standard model Lagrangian density, and do not augment
the field content of the theory.  
Here $\Phi$ is the Higgs field defined
in~(\ref{higgsmatrix}), $F_{\mu\nu}$ is the $SU(2)$ field strength
tensor defined in~(\ref{fieldstrength}) and 
\be
\tilde{F}^{\mu\nu} = \frac{1}{2} \epsilon^{\mu\nu\alpha\beta}F_{\alpha\beta}
\ .
\label{Fdual}
\ee
is the dual of the field strength tensor.
${\cal O}$ is the lowest dimension \CP\ odd operator 
which can be constructed from
minimal standard model Higgs and gauge fields.
Standard model \CP\ violation in the CKM matrix 
does induce the term ${\cal O}$ with $M=v$ in the effective
action, but the coefficient
$b$ is thought to be tiny. 
For us, $b$ is a dimensionless coupling measuring the
strength of the \CP\ violating physics at the scale $M$,
above which our effective theory ceases to be valid.
Throughout most of this paper,
$b$ and $M$ occur only in the combination $b/M^2$, and
in this sense we are introducing a single new parameter.

The operator ${\cal O}$ induces electric dipole moments
for the electron and the neutron, and the strongest experimental
constraint on the size of such an operator come from the
fact that such dipole moments have not been observed.
Working to lowest order (one-loop) we find
\begin{equation}
\frac{d_e}{e} = \frac{m_e \sin^2(\theta_W)}{8\pi^2}\frac{b}{M^2}
\ln\left(\frac{M^2 + m_H^2}{m_H^2}\right)\ .
\label{dipole}
\end{equation}
$M^2$ arises in the logarithm without
$b$ because $M$, the scale above which the effective theory
is not valid, is the ultraviolet cutoff for the divergent loop integral.
A result similar to (\ref{dipole}) was obtained in Ref. \cite{zhang}.
Using the experimental limit\cite{commins} $d_e/e < 4 \cdot 10^{-27}
{\rm cm}$ we find
the bound
\begin{equation}
\frac{b}{M^2}\ln\left(\frac{M^2 + m_H^2}{m_H^2}\right) <
\frac{1}{(3 {\rm ~TeV})^2}\ .
\label{bound}
\end{equation}
The experimental limit\cite{ramsey} on
the neutron electric dipole moment $d_n$ is weaker than that
on $d_e$, but because $d_n$ is proportional to the quark mass rather
than to the electron mass, the constraint obtained using
$d_n$ is comparable to (\ref{bound}).
A baryogenesis scenario which relies on \CP\ violation introduced
via the operator ${\cal O}$ must respect the bound (\ref{bound}).

We wish to estimate the BAU generated  during
a first order electroweak phase transition in the minimal
standard model with the addition of the operator ${\cal O}$ 
with a coefficient $b/M^2$ satisfying (\ref{bound}).  Baryogenesis
can occur either ``locally'' (baryons are produced as a result
of ${\cal B}$ violating processes and ${\cal CP}$ violating processes
occuring together near bubble walls) or
``nonlocally'' (baryons are produced as a result of ${\cal CP}$
violation occurring as particles bounce off bubble walls
and ${\cal B}$ violation occuring away from bubble
walls in the high temperature phase). 
In general,
both local and nonlocal baryogenesis will occur
and the BAU will be the sum of that generated by the two processes.
We have nothing to add to the existing 
treatments\cite{CKN2,CKN4,F&S,GS,JPT,CKV,HN}
of nonlocal baryogenesis, also called baryogenesis by charge
transport.  In this paper, we reconsider the
models of baryogenesis first introduced by
Turok and Zadrozny\cite{TZ} and Dine {\it et al.}\cite{DHSS},
and try to estimate the BAU produced locally.
Throughout this paper, we assume that the time
during which the expectation value of the 
Higgs field is changing at any point in space
is short compared to all other timescales in the
problem.  Throughout, we refer to walls satisfying
this criterion as ``thin'', even though they may
meet the criterion either by being thin enough or by
moving fast enough.  It is worth noting that for 
walls which move at a supersonic velocity, 
diffusion from the wall foward into the high
temperature phase, and hence nonlocal baryogenesis,
are not possible.  In the opposite limit to that
we treat, that is if the walls are ``thick'' in
the sense that the Higgs field changes slowly
relative to all other timescales, then the
effects of the operator $\cal O$ can be
treated approximately 
as a chemical potential for baryon number\cite{DHSS},
realizing a possibility first considered in Ref. \cite{CKN1}
and called spontaneous baryogenesis. 
The local
contribution to the BAU in the thick wall limit was
first estimated by Dine {\it et al.}\cite{DHSS}
and has been further analyzed in Refs. \cite{CKN3,GS,JPT}, and is
not affected by the considerations of this paper.

In this paper we examine two semi-analytical 
methods for estimating the efficiency
of local baryogenesis occurring in the standard model with the addition of the
operator ${\cal O}$, assuming a strongly first order phase transition
and thin bubble walls.
In section III, we explore an approach 
pioneered by Turok and Zadrozny\cite{TZ} to estimate the baryon asymmetry
by considering the relaxation of topologically nontrivial field 
configurations produced during the phase transition. 
In contrast with Ref. \cite{TZ}
we study the full Higgs plus gauge dynamics and, because we introduce \CP\ 
violation via the operator ${\cal O}$ rather than using the two Higgs doublet 
model employed in Ref.~\cite{TZ}, we need not and do not add new fields to 
the theory. Turok and Zadrozny considered effects which occur while
a (thick) bubble wall is passing.  We use their method to consider 
physics after the passage of a thin wall.  Unfortunately, we show
that the use of 
a one-parameter family of spherically symmetric configurations
as in Ref.\cite{TZ}
can give qualitatively misleading results,
and demonstrate the difficulty of
obtaining analytic estimates in this picture without doing a full
scale numerical simulation. 
In Section II
we discuss certain properties of gauge
and Higgs field configurations in the standard electroweak theory,
as a prelude to our discussion of baryon number production
in Section III. 
Included in Section II is a demonstration that when the
\C\ and
\Parity\ transformations are defined
to treat configurations with winding properly, the bosonic sector
of the electroweak lagrangian is not \C\ or \Parity\ invariant
but is \CP\ invariant.  In most treatments, \C\ violation arises
because of the $V-A$ nature of the fermionic part of the theory,
so it is amusing to see that even in the bosonic sector, \C\
is not a good symmetry.
In Section IV we turn to a method introduced by 
Dine {\it et al.}\cite{DHSS}.  
We consider configurations
which happen to be near the crest of the ridge between
vacua as the wall arrives, and estimate the extent 
to which their velocity in configuration space is modified
by the operator ${\cal O}$ during
the passage of the wall.
If the wall is thick, it turns out that the velocity of motion
in configuration space
is {\it not} affected.  The asymmetry arises because ${\cal O}$
affects the potential energy surface in configuration space
during the passage of the wall.  If the wall is thin, as
we assume in this paper, there is no significant time during
which the potential energy surface is affected, but the configuration
space velocities {\it are} affected asymmetrically.
We show how to 
estimate the quantities entering the final result in the thin wall limit
correctly.
Nevertheless, we argue that difficulties 
of the kind encountered in 
Section III also apply to the method of Section IV, rendering the estimate
for the BAU 
more of an upper bound than an estimate. 
Thus, it
seems to us that neither approach can be pushed far enough to obtain
a convincing semi-analytical estimate of the BAU produced by local electroweak
baryogenesis in the thin wall limit, and a large scale numerical
simulation is called for.

\section{Relevant Properties of the Standard Electroweak Theory}

In this section we examine certain properties of the standard model which 
are important for the arguments
of Section III. 
Although our motivation 
is to set the stage for
the next section, we have also endeavoured to write this
section in such a way that it is independent of the
rest of this paper.
The relevant dynamics take place in the bosonic sector.
We consider fermion production in the 
background of the evolving Higgs and gauge fields. We do not take into 
account the back-reaction of the fermions on the bosonic fields. Thus, we 
begin by looking at the purely bosonic part of the standard model where, for 
simplicity, we ignore the $U(1)$ hypercharge gauge field.

\be
{\cal L}=
-\frac{1}{2}\hbox{Tr}(F_{\mu\nu}F^{\mu\nu})
-\frac{1}{2}\hbox{Tr}(D^{\mu}\Phi)^{\dagger}D_{\mu}\Phi
-\frac{\lambda}{4}\left[\hbox{Tr}(\Phi^{\dagger}\Phi)-v^2\right]^2 \ ,
\label{SM action}
\ee
where

\bea
F_{\mu\nu} & = & \partial_{\mu}A_{\nu}-\partial_{\nu}A_{\mu}
-ig[A_{\mu},A_{\nu}] \nonumber \\
D_{\mu}\Phi & = & (\partial_{\mu}-igA_{\mu})\Phi
\label{fieldstrength}
\eea
with $A_{\mu}=A^a_{\mu}\tau^a/2$ where $\tau^a$ are the three
Pauli matrices.
The standard Higgs doublet 
$\vp=(\vp_1,\vp_2)$ is related to the matrix $\Phi$ by

\be
\Phi({\bf x},t)=\left(\matrix{\vp_2^* & \vp_1 \cr
-\vp_1^* & \vp_2} \right) \ .
\label{higgsmatrix}
\ee
Here $v=247$GeV and $g=0.65$. 
The gauge boson mass is $m_W=\frac{1}{2}gv$ and the Higgs
boson mass is $m_H=\sqrt{2\lambda}v$.

Note that

\be
\Phi^{\dagger}\Phi = (\vp_1^*\vp_1 + \vp_2^*\vp_2)\ 
\left(\matrix{1 & 0 \cr 0 & 1} \right) \ ,
\ee
so that we can write

\be
\Phi = \frac{\sigma}{\sqrt{2}}\, U \ ,
\label{sigmaU}
\ee
where $\sigma^2 = 2\left(\vp_1^*\vp_1 + \vp_2^*\vp_2 \right) =
{\rm Tr} \Phi^\dagger \Phi$,  and $U$ 
is an $SU(2)$
valued field which is uniquely defined at any spacetime point where
$\sigma$ does not vanish. Without loss of generality we impose the 
condition that at all times

\be
\lim_{|{\bf x}|\rightarrow \infty} \sigma({\bf x},t) =  v \ ,
\label{sigbc}
\ee

\be
\lim_{|{\bf x}|\rightarrow \infty} U({\bf x},t) =  
\left(\matrix{1 & 0 \cr 0 & 1}\right) \ .
\label{Ubc}
\ee
In $A_0=0$ gauge,
a vacuum configuration is of the form

\bea
\Phi & = & \frac{v}{\sqrt{2}}\,U \nonumber \\
A_j & = & \frac{1}{ig}\partial_jUU^{\dagger} \ .
\label{vacuum}
\eea
At any time $t$ when $\sigma({\bf x},t) \neq 0$ for all ${\bf x}$ we have that
$U({\bf x},t)$ is a map from ${\bf R}^3$ with the points at infinity
identified, that is $S^3$, into $SU(2)$ and therefore $U({\bf x},t)$
can be associated with an integer-valued winding

\be
N_H(t) = w[U]=\frac{1}{24\pi^2} \int d^3x\, \epsilon^{ijk} \hbox{Tr}
[U^{\dagger}\partial_iUU^{\dagger}\partial_jUU^{\dagger}\partial_kU] \ ,
\label{higgswinding}
\ee
the Higgs winding number. 
If $\Phi({\bf x},t)$ evolves continuously
in $t$ then $N_H(t)$ can change only  at times when
there is a zero of $\sigma$ at some point in space. 
At such times, $N_H$ is not defined; at all other times,
it is integer-valued. 
Note that the Higgs winding number of a vacuum configuration (\ref{vacuum})
is equal to its Chern-Simons number

\begin{equation}
N_{CS}(t) = \frac{g^2}{32 \pi^2}\int d^3x\, \epsilon^{ijk}
{\rm Tr}\left( A_i \partial_j A_k + \frac{2}{3}ig A_i A_j A_k \right)\ .
\label{NCSdef}
\end{equation}
For a general non-vacuum configuration the Chern-Simons
number is not integer-valued.

\subsection{Topologically Interesting Configurations}

In this section we are interested in the dynamics of nonzero energy 
configurations with nonzero Higgs winding. A simple example is

\bea
\Phi({\bf x}) & = & \frac{v}{\sqrt{2}}U_{[1]}({\bf x}) \nonumber \\
A_\mu({\bf x}) & = & 0 \ ,
\label{winding1}
\eea
where $U_{[1]}({\bf x})$ is a winding number one map, say,

\be
U_{[1]}({\bf x}) = \exp\left(i\eta(r)\mbox{\boldmath $\tau$}
\cdot\hat{\bf x}\right) \ ,
\label{u1}
\ee
with $\eta(0)=-\pi$ and $\eta(\infty)=0$. The 
configuration~(\ref{winding1}) has no
potential energy but does carry gradient energy because the covariant
derivatives $D_i\Phi$ do not vanish.  This configuration
has $N_H=1$.
If the configuration~(\ref{winding1}) were
released from rest it would radiate away its energy and relax 
towards a vacuum 
configuration. There are two very different ways for this to 
occur\cite{TZ}. If the
characteristic size of $U_{[1]}$ is large 
compared to $m_W^{-1}$,
then the gauge field will evolve
until it lines up with the Higgs field making the covariant derivatives zero,
and at late times $N_H$ will still be one.
If the characteristic size is small the configuration 
will shrink,
the Higgs field $\sigma$ will go through a zero, and at late times $N_H$
will be zero. This dynamics is the subject of
the next section.

Note that $N_H$ is not invariant under large gauge transformations. However,
the change in Higgs winding, $\Delta N_H$, is gauge invariant and the two
distinct relaxation processes are distinguished by whether $\Delta N_H$ is
zero or nonzero. Throughout this section we choose the gauge such that our
prototypical initial configuration is of the form~(\ref{winding1}) which has
$N_H=1$.

The configuration~(\ref{winding1}) 
is similar to the Skyrmion which is a winding
number one soliton in the nonlinear sigma model associated with QCD. The
Skyrme lagrangian has a four derivative interaction which is not present
in~(\ref{SM action}). This term stabilizes the soliton because it prevents the
winding number one configurations from shrinking to zero size. Note that in the
Skyrme model winding number one solitons are identified with baryons and 
winding number minus one solitons are viewed as antibaryons.

We could view the action~(\ref{SM action}) as an effective theory which
describes the low energy degrees of freedom of some more fundamental
theory such as technicolor. The sigma model of~(\ref{SM action}) would
be to technicolor what the usual sigma model which describes pions is to QCD.
The addition of a sufficiently large Skyrme-like term would result in 
classically stable electroweak solitons which we could call techni-skyrmions.
The Higgs winding of such an electroweak soliton would be identified with its 
technibaryon number. 
In this paper we are not modifying~(\ref{SM action}) except for the 
addition of the dimension six \CP\ violating 
term (\ref{newop}) and all configurations with
winding are unstable. 

\subsection{Fermion Production}
A wound up configuration of the form~(\ref{winding1}) is not stable and if
released from rest it will evolve to a vacuum configuration of the 
form~(\ref{vacuum}) plus radiation. In the process fermions may be 
anomalously produced. If the fields relax to the vacuum by changing the Higgs 
winding then there is no anomalous fermion number production. However, if 
there is no net change in 
Higgs winding during the evolution (for example $\sigma$ never
vanishes) then there is anomalous fermion number production.

To understand these claims consider two sequences of configurations beginning 
with the wound up 
configuration~(\ref{winding1}) and ending at the classical 
vacuum~(\ref{vacuum}). The first sequence ends at the vacuum~(\ref{vacuum})
with $U={\bf 1}$ while the second ends up at $U=U_{[1]}$.
Note that these sequences cannot be
solutions to the classical equations of motion since the initial
configurations carry energy whereas the final ones do not.
Throughout both
sequences we maintain the boundary conditions~(\ref{sigbc}) and~(\ref{Ubc}).
For the first sequence, $\sigma$ must vanish at some intermediate 
configuration since the Higgs winding changes. For the second sequence,
the change in Higgs winding is zero and $\sigma$ need not vanish.

Now introduce an $SU(2)_L$ weak fermionic doublet, $\psi$. The fermion is 
given mass through the usual gauge invariant coupling to the Higgs field 
$\Phi$ and for simplicity we assume that both the up and down components of 
the doublet have the same mass, $m$. The fermion field is quantized in the 
background of the bosonic fields given by our interpolation. 

Now, the anomaly equation

\be
\partial_{\mu}J^{\mu} = \frac{g^2}{32\pi^2}\hbox{Tr}(F{\tilde F}) \ ,
\label{anomaly}
\ee
when integrated, implies that the change in the fermion number from the
beginning to the end of a sequence is given by

\be
\left.\int d^3x\, J^0\right|_{\rm final} - 
\left.\int d^3x\, J^0\right|_{\rm initial} = -w[U] \ ,
\label{intanomaly}
\ee
where $U$ is that of the final configuration~(\ref{vacuum}). For the
first sequence $w$ is one whereas for the second it is zero. 
Thus fermion number is violated in processes for which the 
configuration~(\ref{winding1}) 
unwinds via gauge unwinding, but is not violated when 
such a configuration unwinds via a Higgs unwinding.

For both of the interpolations which we have considered, the final 
background configuration is a vacuum configuration of the form~(\ref{vacuum}).
In this background the lowest energy fermion state has fermion number zero and
the fermion number of any other state with $n$ $\psi$ particles and $m$ 
$\psi$ antiparticles is $n-m$. However, the fermion number of the initial 
state, where the background has the non-vacuum form~(\ref{winding1}), is more
complicated. Suppose that the winding number one map $U_{[1]}$ 
in~(\ref{winding1}) 
has a characteristic size $L$. It is known that if $mL \gg 1$
then the lowest energy fermion state in the presence of this background has
fermion number one\cite{goldwil,dhfar,mackwil}. 
In this sense the configuration~(\ref{winding1}) is said to
carry fermion number. If our initial fermion state is this lowest energy
state and we let this state evolve in the changing background of our first
interpolation then the final state will contain one net $\psi$ particle.
Although a $\psi$ particle must be produced, there is no violation of 
fermion number since both initial and final states have fermion number one.
If we evolve the same initial state in the background of the second
interpolation there will be no net $\psi$ particles in the final state.
This is consistent with anomalous fermion production as described by
equation~(\ref{intanomaly}). On the other hand, if we begin with the 
configuration~(\ref{winding1}) and $mL \ll 1$, then the lowest energy state in
this background has fermion number zero. If this state evolves in the
background of the first interpolation then there are no net $\psi$ particles
in the final state whereas if it evolves in the background of the second
interpolation there is one net antifermion produced.
This discussion of where the fermion number resides does not alter the
general conclusion which is that if we smoothly 
interpolate from~(\ref{winding1})
to~(\ref{vacuum}) with $U={\bf 1}$ then there is no anomalous fermion 
production in this background whereas if we end up at~(\ref{vacuum}) with
$U=U_{[1]}$ then there is anomalous fermion production.

Of course, we are actually interested in the dynamical evolution of
configurations such as~(\ref{winding1}) 
which are released from rest and end up as
outgoing radiation. In this case it is dangerous to use the anomaly
equation~(\ref{anomaly}) since $\int d^4x\,\hbox{Tr}(F{\tilde F})$ is not
well-defined as an integral and any answer can be obtained for the change in 
fermion number\cite{fiveofus}. Nonetheless, the results of our previous
discussion still apply. If the configuration~(\ref{winding1}) 
is released and falls
apart without ever going through a zero of the Higgs field, then the 
analysis of Ref. \cite{FGLR} is directly applicable and we conclude that
one net antifermion is produced just as we did with the second of our
interpolations.  
If the Higgs field unwinds by going through a zero, then
one can use arguments presented in 
Refs. \cite{fiveofus,FGLR}
to demonstrate that
the presence of outgoing
radiation in the final configuration does not affect
the result above, namely that there is no fermion
number violation.

\subsection{Discrete Symmetries: \C, \Parity\ and \CP}
In this subsection we study the properties of the 
lagrangian~(\ref{SM action}) under the discrete transformations \C, \Parity\ 
and \CP. We must keep in mind that certain configurations of $\Phi$ can be
associated with a particle number which equals the Higgs winding defined
by~(\ref{higgswinding}) and~(\ref{sigmaU}). With the identification of
winding with particle number, we will see that~(\ref{SM action}) is not
invariant under \C\ and \Parity\ separately but is invariant under \CP.

First let us describe the parity transformation. Let \Parity$'$ be the
obvious discrete transformation defined as

\bea
{\cal P}' & : & A_0({\bf x},t) \rightarrow A_0(-{\bf x},t) \nonumber \\
{\cal P}' & : & A_i({\bf x},t) \rightarrow -A_i(-{\bf x},t) \nonumber \\
{\cal P}' & : & \Phi({\bf x},t) \rightarrow \Phi(-{\bf x},t) \ .
\eea
Under the parity transformation \Parity$'$\ the lagrangian~(\ref{SM action})
is invariant. However, under \Parity$'$\ the winding number given 
by~(\ref{higgswinding}) changes sign and we certainly do not want
particle number to change sign under parity! Now parametrize the unitary
matrix $U$ appearing in~(\ref{sigmaU}) by the fields 
$\pi^a({\bf x},t)$ which are weak scale sigma model analogs of the low
energy pions

\be
\Phi({\bf x},t) = \frac{\sigma({\bf x},t)}{\sqrt{2}}\exp\left(
i\tau^a\pi^a({\bf x},t)\right) \ .
\label{pions}
\ee
We can define a parity operator \Parity\ which takes
$\pi^a({\bf x},t)$ to $-\pi^a(-{\bf x},t)$ which is the conventional
transformation property of the ordinary pions. Consistent with this we
define

\bea
{\cal P} & : & A_0({\bf x},t) \rightarrow A_0(-{\bf x},t) \nonumber \\
{\cal P} & : & A_i({\bf x},t) \rightarrow -A_i(-{\bf x},t) \nonumber \\
{\cal P} & : & \Phi({\bf x},t) \rightarrow \Phi^{\dagger}(-{\bf x},t) \ .
\label{trueP}
\eea
With this definition of parity, which we adopt, the winding $w[U]$ is
unchanged by a parity transformation. However, in this case the 
weak interaction lagrangian~(\ref{SM action}) is not parity invariant.

Now turn to charge conjugation. If we include the $U(1)$ interaction we
see that $A_{\mu}^3$ is a linear combination of the photon and the
$Z$-boson. Since the photon is charge conjugation odd we certainly
want $A_{\mu}^3 \rightarrow -A_{\mu}^3$. Similarly we want
$W_{\mu}^+ \rightarrow -W_{\mu}^-$ and $W_{\mu}^- \rightarrow -W_{\mu}^+$
which is the same as $A_{\mu}^1 \rightarrow -A_{\mu}^1$ and
$A_{\mu}^2 \rightarrow A_{\mu}^2$. This is equivalent to the requirement 
that the matrix $A_{\mu}=A_{\mu}^a\tau^a/2$ transforms into
$\tau_2A_{\mu}\tau_2$. Thus we can attempt to define charge conjugation
by \C$'$\ where

\bea
{\cal C}' & : & A_{\mu} \rightarrow \tau_2A_{\mu}\tau_2 \nonumber \\
{\cal C}' & : & \Phi \rightarrow \tau_2\Phi\tau_2
\label{falseC}
\eea
This transformation has the property that $\vp_1 \rightarrow \vp_1^*$ and
$\vp_2 \rightarrow \vp_2^*$ which are the expected transformations of the 
complex fields and furthermore leaves the lagrangian~(\ref{SM action})
invariant. 
Note that the Lagrangian~(\ref{SM action}) is invariant
under the transformation $\Phi \rightarrow B^{\dagger}\Phi B$,
$A_{\mu}\rightarrow B^{\dagger}A_{\mu}B$, where $B$ is a spacetime
independent $SU(2)$ matrix. With $B=i\tau_2$ we obtain the
transformation~(\ref{falseC}). 
Thus, the transformation \C$'$\ is one element of a continuous
global symmetry group and should not be viewed as a discrete
transformation. 
In addition, under \C$'$\ the winding number
is unchanged which is unacceptable.

Return to the parametrization of $\Phi$ given by~(\ref{pions}). Under 
charge conjugation we expect $(\pi^1,\pi^2,\pi^3) \rightarrow
(\pi^1,-\pi^2,\pi^3)$. To be consistent with this let us define

\bea
{\cal C} & : & A_{\mu} \rightarrow \tau_2A_{\mu}\tau_2 \nonumber \\
{\cal C} & : & \Phi \rightarrow \tau_2\Phi^{\dagger}\tau_2
\label{trueC}
\eea
Adopting this definition we have that the winding flips sign under
charge conjugation. However, the weak interaction 
lagrangian~(\ref{SM action}) is not \C\ invariant.

Consider \CP\ defined as the composition of~(\ref{trueP}) 
and~(\ref{trueC}):

\bea
{\cal CP} & : & A_0({\bf x},t) \rightarrow \tau_2A_0(-{\bf x},t)\tau_2 
\nonumber \\
{\cal CP} & : & A_i({\bf x},t) \rightarrow -\tau_2A_i(-{\bf x},t)\tau_2 
\nonumber \\
{\cal CP} & : & \Phi({\bf x},t) \rightarrow \tau_2\Phi(-{\bf x},t)
\tau_2 \ .
\label{trueCP}
\eea
All fields transform as expected, for example $\vp_1({\bf x},t)
\rightarrow \vp_1^*(-{\bf x},t)$, the lagrangian~(\ref{SM action}) is
invariant under \CP\ and the Higgs winding changes sign. 
Note that ${\cal CP} = {\cal C}'{\cal P}'$.
In defining
\C\ and \Parity\ one must be careful to ensure that
winding flips under \C\ but not under \Parity .  In so doing, 
one discovers that
with appropriate definitions of \C\ and \Parity\ the {\it bosonic} 
sector of the 
electroweak lagrangian is not \C\ or \Parity\ invariant but is 
\CP\ invariant. 

\section{Local Baryogenesis Through Unwinding}

In this section we explore the possibility of
using the method of Turok and Zadrozny\cite{TZ}
to estimate the baryon asymmetry produced by local baryogenesis
in a scenario in which the electroweak phase transition
is strongly first order and the bubble walls are thin.
The theory we treat is the standard model augmented
by the \CP\ violating operator ${\cal O}$ of (\ref{newop}). 
Turok and Zadrozny\cite{TZ} studied the classical dynamics of 
topologically
nontrivial gauge and Higgs field 
configurations in the presence of \CP\ violation. 
Following Ref. \cite{TZ}, we begin by considering
spherically symmetric nonvacuum configurations of the form 
(\ref{winding1}) with Higgs winding $N_H=\pm 1$
and discuss their dynamics when they are released from rest and 
evolve according
to the equations of motion.  Solutions to the equations of motion
typically approach a vacuum configuration uniformly throughout
space at late times, and these solutions are no exception.
There are, however, two qualitatively 
different
possible outcomes of this evolution.
One possibility is that at some time during the evolution,
$\sigma$, the magnitude of the Higgs field, is zero
at some point in space and at late times 
the configuration
tends toward a vacuum with $N_H=0$.
As we have seen in Section II, fermion number is not
violated if the configuration unwinds in this
fashion.  The second possibility is that at late times the configuration
dissipates toward a vacuum with $N_H$ unchanged from
its initial value of $\pm 1$.  In other words,
the Higgs field does not unwind but instead 
the gauge field ``winds up''. 
We saw in Section II that fermions are produced in the background
of a solution which evolves in this way.
Without \CP\ violation, for every $N_H=+1$
configuration which relaxes in a baryon producing fashion
there is an $N_H=-1$ configuration which produces anti-baryons.
With the inclusion of the \CP\ violating
operator ${\cal O}$, the hope is that there will 
be some configurations which produce baryons whose \CP\
conjugate configurations relax to the $N_H=0$ vacuum without
violating baryon number.

We now pause to say a few words about the dynamical context
in which we wish to use the scenario just described.
We imagine that the (thin) bubble wall has just passed,
leaving in its wake the
configuration we are looking at, but that this configuration
has not yet had time to relax to equilibrium.  Our goal is a 
qualitative understanding of the dynamics of this relaxation.
A first order 
electroweak phase 
transition
can be characterized by the change in the gauge
invariant quantity $\langle \sigma^2 \rangle$.
If we renormalize $\langle \sigma^2 \rangle$ such that it
is equal to $v^2$ at zero temperature, then for a strongly
first order phase transition it is close to
$v^2$ just below $T_c$ in the low temperature
phase, and is much smaller 
just above $T_c$ in the high temperature phase.
This is a slight motivation for considering initial configurations
in which $\sigma=v$ throughout space, 
even though this is  in reality not 
a good description of the non-equilibrium configurations left in the
wake of the wall 
and is in fact
not maintained during the subsequent evolution.
There is no justification for choosing either a spherically
symmetric configuration, or one with $A_\mu=0$, or one which
is initially at rest.  We shall see that this is the
Achilles heel of the whole approach.

It is worth noting that Turok and Zadrozny introduced the
analysis we are describing in a somewhat different context.
They considered the evolution of configurations during
the passage of a thick wall, rather than after the passage
of a thin wall.  Instead of using the operator (\ref{newop})
to introduce \CP\ violation, they began with the
two Higgs doublet model and used the 
operator 
\begin{equation}
\theta {\rm Tr} F\tilde F
\label{TZop}
\end{equation}
where $\theta$ is a \CP\ odd phase between the two Higgs fields.
(Although the operator (\ref{TZop}) is \CP\ even, \CP\ violating
dynamics in the Higgs sector picks a sign for the change in
$\langle \theta \rangle$ during the phase transition, thus
communicating the \CP\ violation to the gauge sector.)
To avoid the complications of
simulating all the fields in the two Higgs doublet model, 
they simplified the problem as follows.
First, they considered
circumstances in which $\langle \theta \rangle$
was spatially uniform but changes
linearly in time during the passage of the bubble wall.
That is, they replaced $\theta({\bf x},t)$ in (\ref{TZop}) by 
$\theta(t) = ct$ with $c$ a constant.  Second, they chose to consider
the effect of this forcing term, now treated as externally
imposed, on just the standard model fields.
That is, they found 
solutions to the equations of motion obtained 
from a
Lagrangian given by (\ref{SM action}) with the addition of the term
$(ct\,{\rm Tr}F\tilde F)$.  Since this Lagrangian
depends explicitly on time, there is no conserved
energy in their problem. 
If we wanted to explore the effects of 
the \CP\ violating
operator (\ref{newop}) used in this paper as $\langle \sigma^2 \rangle$
changes during the passage of the wall, we could
follow precisely the same strategy.
Instead, we wish to treat the non-equilibrium 
conditions after the passage of a thin wall.
Therefore, we simply use a Lagrangian given by (\ref{SM action})
plus (\ref{newop}).  
This means that 
energy 
is conserved (to better than half a percent in our numerical
simulations) during the evolution of the gauge and Higgs fields.

We wish to solve the equations of motion obtained from
the action (\ref{SM action}) augmented by the addition
of the operator (\ref{newop}).  We work in the spherical
ansatz\cite{ansatz} in which all gauge invariant quantities
are functions only of $r$ and $t$, and solve the equations
numerically.  In this paper, we 
do not describe our numerical methods in any detail; they are standard
and are similar to those of Ref. \cite{FGLR}.
We also do not present our solutions 
in full. Rather, we describe them only in sufficient
detail to demonstrate that it is very difficult
to follow this approach to completion.
Let us first consider initial conditions 
of the form (\ref{winding1}) with $\eta(r)$ of (\ref{u1})
given by 
\begin{equation}
\eta(r) = -\pi \left[ 1- {\rm tanh}\left(\frac{r}{R}\right) \right]
\label{eta}
\end{equation}
where $R$ is a constant parametrizing the size of the configuration.
This configuration satisfies the boundary condition (\ref{Ubc})
and has Higgs winding number $N_H=+1$, and we use it 
as the initial condition for the 
equations of motion, setting all time derivatives to zero at $t=0$.
For the moment, we set
$b=0$ in (\ref{newop}) and do not introduce \CP\ violation.
In agreement with Turok and Zadrozny, we find that 
there is a critical value of $R$, which we call $R_c^+$, defined
as follows.  For all $R<R_c^+$ the configuration 
evolves toward a vacuum configuration with $N_H=0$.  No fermions
are produced in this background.  For all $R>R_c^+$, the configuration
evolves toward a vacuum configuration with $N_H=+1$, and
fermions are produced.  The values of $R_c^+$ which we obtain
in simulations with several different values of $m_H/m_W$ 
are in quantitative agreement with those obtained in Ref. \cite{TZ}.
Repeating this exercise beginning
with $\eta(r)$ given by $-1$ times that in (\ref{eta}), that
is beginning with the \CP\ conjugate configuration having
$N_H=-1$, we find an analogously defined $R_c^-$.  As we
have not yet introduced any \CP\ violation, 
we necessarily find $R_c^-=R_c^+$.
We now repeat the entire procedure with $b\neq 0$, that is with \CP\
violation present.  
Although at this point quantitative comparison with Ref. \cite{TZ}
is no longer possible because of the differences described
above,  
we do in fact find $R_c^-\neq R_c^+$. Unfortunately, this is not the end
of the story.

We now investigate slightly more general initial conditions. 
Consider initial configurations exactly as above
except that the time derivative of $\sigma$ is nonzero and
is given by
\begin{equation}
\dot\sigma(r) = \gamma v^2 \left[ 1 - {\rm tanh}\left(r/R\right)\right]\ ,
\label{sigmadot}
\end{equation}
with $\gamma$ some constant.  We now find that for 
some values of $\gamma$, 
$R_c^- < R_c^+$ whereas for other
values of $\gamma$, $R_c^- > R_c^+$.
This dooms an analysis in terms of the single parameter $R$.
Clearly, a more general framework is needed.

Consider a family of initial configurations with $N_H=+1$, much more
general than we have considered to this point, parametrized
by a set of parameters ${\beta_i}$.  To this point, we
have introduced two such parameters, $R$ of (\ref{eta}) and (\ref{sigmadot})
and $\gamma$ of (\ref{sigmadot}).  More generally, we must
consider other profile functions for $\eta$ and $\dot\sigma$ and
must allow for initial 
configurations in which  $\sigma$ varies with $r$ and $A_\mu$ and
$\dot A_\mu$ are nonzero.  More generally still, one would
want to go beyond the spherical ansatz.
Of course eventually one wants to 
work towards an analysis involving an infinite set of $\beta$'s,
but it seems reasonable to start with some finite set ${\beta_i}$.
For any fixed $b$, define a function $F^+(\beta_1,\beta_2,\ldots)$ 
which has the following properties.  
$F^+(\beta_1,\beta_2,\ldots)>0$ for all points in $\beta$-space
which describe configurations which evolve toward the $N_H=+1$
vacuum thereby producing fermions, 
and  $F^+(\beta_1,\beta_2,\ldots)<0$ for all 
points in $\beta$-space describing
configurations which evolve towards the $N_H=0$ vacuum.
If we only consider  
initial configurations described by (\ref{winding1}),
(\ref{u1}) and (\ref{eta}) which are parametrized by the
single parameter $R$, 
then we can take $F^+(R) = R-R_c^+$. 
Completely analogously, we define a function $F^-(\beta_1,\beta_2,\ldots)$
such that the hypersurface $F^-(\beta_1,\beta_2,\ldots)=0$ 
divides the $\beta$-space
of $N_H=-1$ configurations into those which evolve towards the 
$N_H=-1$ and $N_H=0$ vacua.

In the absence of \CP\ violation, 
$F^+(\beta_1,\beta_2,\ldots)=0$ and $F^-(\beta_1,\beta_2,\ldots)=0$ define
the same hypersurface. In this case, imagine allowing 
a \CP\ symmetric ensemble 
of configurations with $N_H=+1$ and $N_H=-1$ to evolve.
(By \CP\ symmetric we mean that the probability for finding
a particular $N_H=+1$ configuration in the ensemble
is equal to that for finding its \CP\ conjugate $N_H=-1$ configuration.)
Since the configurations which anomalously produce fermions
are exactly balanced by those which anomalously produce antifermions, 
the net fermion number produced after relaxation would be zero. 
We wish to investigate the behavior in the 
presence of the \CP\ violating term ${\cal O}$ of
(\ref{newop}). The hope is that ${\cal O}$ will 
affect the dynamics of $N_H=+1$ configurations 
and $N_H=-1$ configurations
in qualitatively different ways
and that after relaxation to vacuum a net
fermion number will result, even though the initial 
ensemble of configurations
was \CP\ symmetric.  
With $b\neq 0$ we have seen that the two hypersurfaces
$F^+=0$ and $F^-=0$ are indeed distinct. 
The configurations represented by points in $\beta$-space 
between the two hypersurfaces yield a net
baryon asymmetry. There are two qualitatively
different possibilities, however, which we illustrate schematically
in Figure 1.  
\begin{figure}
\centerline{
\epsfbox{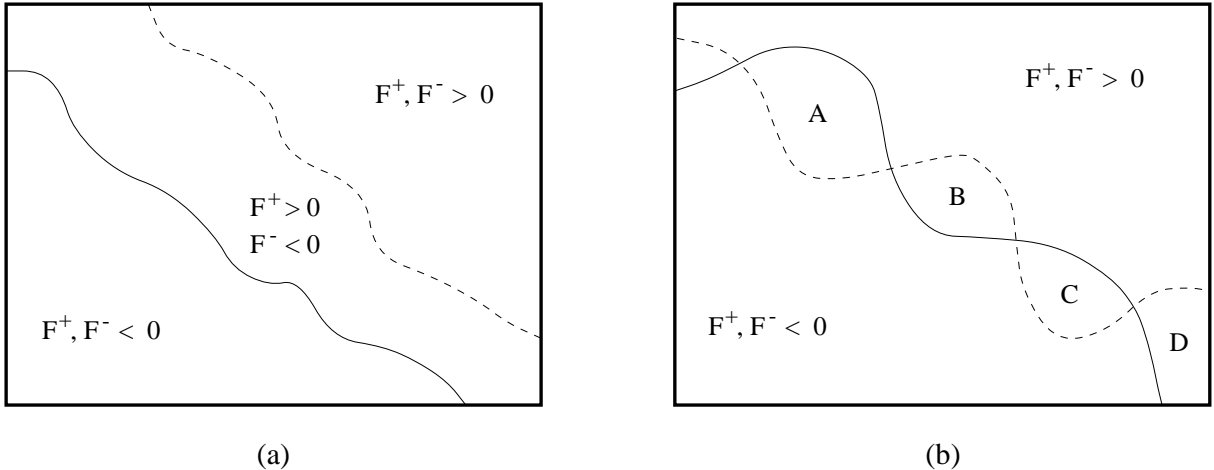}
}
\vspace{0.3in}
\caption{Sketch of two qualitatively different
possible behaviors of the critical surfaces
$F^+=0$ (represented by a solid line) and $F^-=0$ 
(represented by a dashed line) in the space $(\beta_1,\beta_2,\ldots)$
describing initial configurations.}
\label{surf}
\end{figure}
In Figure 1a, the hypersurfaces $F^+=0$ and $F^-=0$
do not cross. The sign of $b$ has been chosen such that
we produce net baryons in the region between the
two surfaces.
In Figure 1b, the hypersurfaces cross and
we produce net baryons in regions $B$ and $D$,
and net antibaryons in regions $A$ and $C$.
If the hypersurfaces do not cross,
as in Figure 1a, then a simple
estimate of the fraction of configurations which yield a 
net baryon asymmetry is possible.  This fraction
would be proportional to the separation between the 
two hypersurfaces measured in any direction in $\beta$-space with
a component perpendicular to the hypersurfaces --- for example, it
would be proportional to
($R_c^+ - R_c^-$) --- and it would be proportional
to $b/M^2$, the coefficient of ${\cal O}$.
Unfortunately, we have seen that when we 
consider the two parameter space of $(R,\gamma)$ the
hypersurfaces $F^+=0$ and $F^-=0$ do in fact cross.
Thus, in the more general space $(\beta_1,\beta_2,\ldots)$
the picture cannot look like that  sketched in Figure 1a
and must look like that sketched in Figure 1b.

There are other indications that life is complicated, as in
Figure 1b.  We have found that the evolution of configurations near the
$F^+=0$ and $F^-=0$ surfaces 
is in some cases extremely complicated.  For example,
there are instances in which just to one side of $F^+=0$, 
the configuration changes its winding number from $N_H=1$ to 
$N_H=0$ to $N_H=1$ to $N_H=0$ to $N_H=1$, having a zero
of the Higgs field at four different times, before finally
settling toward the $N_H=1$ vacuum, whereas just on the
other side of $F^+=0$ the configuration goes through three
zeroes of the Higgs field before settling towards the $N_H=0$ vacuum.
This behavior, also seen by Turok and Zadrozny, 
suggests that the location of the $F^+=0$ surface in $\beta$-space
depends on details of a complicated nonlinear evolution, and
so must be a highly nontrivial function of many of the parameters
specifying the initial configuration.  
It would be very nice to find a simple characterization of
the $F^+=0$ and $F^-=0$ surfaces in terms of only one or a few
parameters, but this seems very unlikely.
Hence, although the dynamics of the unwinding 
of topological configurations after the phase transition
in the presence of the operator ${\cal O}$ may 
lead to a baryon asymmetry,
we see no way to make a simple analytical
estimate of this asymmetry.
This may well be a valid way of looking at the microphysics
of electroweak baryogenesis, but it seems that large scale
$3+1$ dimensional numerical
simulations of the kind recently pioneered by 
Moore and Turok\cite{MT}
(but including \CP\ violation via (\ref{newop}) and working
in a setting in which the bubble walls are thin and rapidly moving)
are required in order to estimate the contribution to the BAU.
In Section V, we will return to a brief discussion of the
large scale numerical simulations which seem necessary.
Before that, in the next section, we explore a different
attempt at obtaining a semi-analytical estimate of the magnitude
of the effect.

\section{Kicking Configurations Across the Barrier}

Let us now turn to what may seem 
initially to be a somewhat orthogonal discussion of the
physics of local electroweak baryogenesis, following that
of Dine, Huet, Singleton, and Susskind\cite{DHSS}. In a sense,
this discussion is more general than that of the previous
section, because it attempts to treat baryon number violating
processes of a type more general than the unwinding of winding number
one configurations.  On the other hand, the treatment
of these more general processes is, of necessity, greatly
over-simplified.  The complications we encountered in the previous
section are real, but they seem  not to appear in the treatment
of this section,
and we are able to obtain an estimate for the baryon asymmetry generated. 
{\it A priori} this could either mean that the method extracts the
essence of the matter or that it sweeps important physics under the rug.  
The truth
is somewhere in between these two extremes.
We give a
discussion of the physics under the rug, and argue that the 
estimate we obtain should be seen as an upper bound on the
baryon asymmetry generated by local electroweak baryogenesis
in the thin wall scenario.
Although we follow
Ref. \cite{DHSS} to some extent, our discussion does
not exactly parallel theirs and we shall note the points
where we differ as we come to them. 

In the high temperature phase, baryon number violating 
processes are not exponentially suppressed.
The barrier crossing configurations typically\cite{AM,ASY}
have sizes given by the magnetic correlation length
\begin{equation}
\xi \sim (\alpha_W T)^{-1}\ .
\label{magcor}
\end{equation}
We will think of dividing space up into cells of this size,
and looking at configurations cell by cell.
The energy in gauge field oscillations with wavelength $\xi$
is of order $T$, but the total energy in a cell is much larger,
as it is presumably of order $T^4\xi^3$. Indeed, this energy is
much larger than the sphaleron energy, which is 
$E_{\rm sph}\sim v/g$.  Most of
the energy is in oscillations of the gauge and Higgs fields on
length scales shorter than $\xi$. These configurations
are crossing the barrier between vacua via regions of the barrier
far above the lowest point on the barrier, that is far above the
sphaleron, and they look nothing like the sphaleron. 
It was initially thought\cite{AM} that in each cell of volume $\xi^3$,
the sphaleron barrier was crossed once per time $\xi$, leading
to a baryon number violation rate 
per unit volume of the form (\ref{kappa})
with $\kappa\sim 1$, in agreement with numerical
simulations\cite{ambjorn}.  A recent analysis\cite{ASY} in fact suggests that
the time it takes for a configuration in a given cell to cross the
barrier is of order $\xi/\alpha_W$ leading to $\kappa\sim \alpha_W$.
We now consider what happens when the bubble wall hits
the configurations just described.

Consider the configuration in one cell.  It traverses a path
through configuration space, which we parametrize by $\tau$.
Dine {\it et al.} consider the special case in which this
path is the path 
in configuration space which an instanton follows as a function of
Euclidean time $\tau$, but this is not essential, and it is clear that they
were thinking of more general circumstances also.
The configurations discussed in Section III can be seen as
special cases of those described here.
The energy of the configuration has
a maximum at some $\tau$ (at which the configuration crosses
the barrier) which we define to be $\tau =0$.  Following Dine {\it et al.},
we now write down a Lagrangian which is intended to describe the 
dynamics of $\tau$ as a function of time for $\tau$ near $\tau=0$:
\be
{\cal L}(\tau,{\dot \tau})=\frac{c_1}{2\xi}{\dot \tau}^2 +
\frac{c_2}{2\xi^3} \tau^2 + 
\frac{c_3}{\xi} \frac{b}{M^2} \sigma^2\, {\dot \tau} \ .
\label{Leq}
\ee
In this expression, $c_1$, $c_2$, and $c_3$ are dimensionless constants,
different for each of the infinitely many possible 
barrier crossing trajectories.
The factors of $\xi$ have been put in by dimensional analysis treating
$\tau$ as a quantity of dimension $-1$. (We
shall see that rescaling $\tau$ by a dimensionful
constant does not change the final result.)
This Lagrangian should be seen
as the first few terms in an expansion in powers of $\tau$ and
$\dot \tau$.  Because we have assumed that $\tau =0$ is a maximum
of the energy as a function of $\tau$, no odd powers of $\tau$
can appear.  
In the absence of \CP\ violation, there can be
no odd powers of $\dot\tau$, since they make the dynamics 
for crossing the barrier from left to right different than from right
to left.  The ${\rm Tr} F\tilde F$ in the operator ${\cal O}$
includes a term which is proportional to the time derivative
of the Chern-Simons number, and this means that ${\cal O}$ must
contribute a term in ${\cal L}$ which is linear in $\dot \tau$.
It is obviously quite an over-simplification to treat barrier
crossing as problem with one degree of freedom.  As we saw
in Section III, the complete dynamics can be very complicated, even
for relatively simple initial conditions.  We will return to 
this point, but for now we forge ahead with (\ref{Leq}).

The momentum conjugate to $\tau$
is given by
\be
p = \frac{c_1}{\xi}{\dot \tau} + \frac{c_3}{\xi} \frac{b}{M^2} \sigma^2
\label{mom}
\ee
and the Hamiltonian density is therefore
\be
{\cal H}= \frac{\xi}{2c_1}\left(p - \frac{c_3}{\xi} 
\frac{b}{M^2} \sigma^2 \right)^2
-\frac{c_2}{2\xi^3} \tau^2 \ .
\label{Heq}
\ee
Before considering the thin wall case of interest in this paper,
it is worth pausing to consider the 
thick wall limit in which $\langle \sigma^2 \rangle$
is changing slowly and other quantities evolve adiabatically
in this slowly changing background. 
A reasonable assumption is 
that the
variables $(\tau,p)$ are Boltzmann distributed with respect to the
Hamiltonian~(\ref{Heq}) at each instant,
treating $\sigma^2$ as approximately constant. 
This implies that the distribution
of $p$ is centered at
\be
p_0 = \frac{c_3}{\xi}\frac{b}{M^2} \sigma^2 \ .
\ee
However, from~(\ref{mom}) we see that this means that the velocity
${\dot \tau}$ is Boltzmann distributed with center ${\dot \tau}= 0$. Thus,
the presence of the \CP\ violating operator~(\ref{newop}) does not bias
the velocity of trajectories in configuration space in the thick wall
limit. This conclusion disagrees with that of Ref. \cite{DHSS}.
There is nevertheless an effect.  Integrating the third term
in (\ref{Leq}) by parts, one obtains a term linear in $\tau$
proportional to the time derivative of $\sigma^2$.  This 
changes the shape of the potential energy surface in configuration
space during the passage of the wall, and yields an asymmetry.
In this limit, in which the wall is thick and departure from
equilibrium is small, the problem is much more easily 
treated in the language of spontaneous 
baryogenesis\cite{CKN1,DHSS,CKN3} ---
the operator ${\cal O}$ acts like a chemical potential for baryon
number. 

We now turn to the thin wall case.  Immediately after the wall strikes,
the fields are not yet in equilibrium.  As we saw in the previous
section, life is complicated. The idea of this section, however,
is to use an impulse approximation to estimate the kick 
which $\dot\tau$ receives as the wall passes, and from this to estimate
the baryon asymmetry that results.  The equation of motion
for $\tau$ obtained from (\ref{Leq}) is
\begin{equation}
\ddot\tau = \frac{c_2}{c_1\xi^2}\tau - \frac{c_3}{c_1} 
\frac{b}{M^2}\frac{d}{dt}\sigma^2 \ .
\label{eofm}
\end{equation}
During the passage of a thin wall, the first term on the right
hand side can be neglected relative to the second.  Making this
impulse approximation, we find that the passage
of the wall kicks  $\dot\tau$ by an amount
\be
\Delta{\dot \tau} = -\frac{c_3}{c_1}\frac{b}{M^2}\Delta \sigma^2 \ ,
\label{deltataudot}
\ee
where $\Delta \sigma^2$ is the amount by which $\sigma^2$
changes at the phase transition.
The kick $\Delta \dot\tau$ has a definite sign.
Thus, in the thin-wall limit, the distribution of the 
velocities in configuration space of 
barrier crossing
trajectories is biased and a baryon asymmetry, whose 
magnitude we now discuss, results.

If $\Delta \dot\tau$ is large compared to $\dot\tau_0$, the velocity
the configuration would have had as it crossed $\tau =0$ in the
absence of the action of the wall, then $\Delta \dot\tau$ 
will kick the configuration
over the barrier in the direction it favors, and will produce, say,
baryons rather than anti-baryons.  If $\Delta \dot\tau$ is small
compared to $\dot\tau_0$, it will have no qualitative effect.  The fraction
of the distribution of configurations 
with $\dot\tau_0 < \Delta\dot\tau$ is proportional
to $\Delta\dot\tau$. (Note that had we taken $\tau$ to have 
dimension other than $-1$ in (\ref{Leq}), various powers of 
$\xi$ would have run through the calculation until this point when
they would have cancelled in computing the {\it fraction} of
the distribution of $\dot\tau$ affected by the wall.)
Note that in this calculation it was not necessary for $\tau$
to be precisely at $\tau=0$ when the wall hits.  It was only necessary
for $\tau$ to be close enough to $\tau=0$ that the Lagrangian
(\ref{Leq}) is a good approximation.   It is difficult to quantify
what fraction of configurations satisfy this criterion
of being ``close enough to $\tau=0$'', so
we simply parametrize our ignorance by calling this
fraction $f$, for fudge factor.  It is worth
noting that $f$ does not depend on the time it takes
configurations to traverse the barrier.  As discussed
at the beginning of the section, this is now thought to be of order
$\xi/\alpha_W$.  Making this time, say, longer just makes
the time during which the configuration is ``close enough to $\tau=0$''
longer and need not affect $f$. Tabling 
further discussion of $f$ momentarily, we estimate 
the net number density of baryons produced 
as
\begin{equation}
n_B \sim \Delta \dot\tau \, f \, \xi^{-3}\ ,
\end{equation}
where we have absorbed the constants $c_i$ into $f$.

At the time of the electroweak phase transition, the entropy
density of the universe is $s\sim 45 T^3$, and so we 
obtain\footnote{This result agrees with that of Ref. \cite{DHSS}, although
our discussion and theirs are somewhat different.  In comparing
to Ref. \cite{DHSS} note that they take $b\sim\alpha_W$.}
\begin{equation}
\frac{n_B}{s} \sim f\, \frac{\alpha_W^3}{45}\,\frac{b}{M^2}\Delta\sigma^2\ .
\label{gettingthere}
\end{equation}
The size of the effect clearly depends on $\Delta \sigma^2$. It has been 
suggested\cite{DHSS} that $\Delta \sigma^2$ 
corresponds to increasing 
$\sigma^2$ up to that value at 
which baryon number violating processes become
exponentially suppressed in thermal equilibrium. 
Whereas in the thick wall case, baryon number violating processes
stop when $\sigma^2$ reaches this value, this is not the 
case in the thin wall scenario.  In this setting, thermal equilibrium
is not maintained even approximately, and we see from the above discussion that
what matters is the net change in $\sigma^2$ as the wall passes.
(If we assume that $\sigma^2$ changes arbitrarily rapidly as we
did by using the impulse approximation above, then the final
answer can only depend on the total jump in $\sigma^2$ and cannot
depend on the value of $\sigma^2$ at which equilibrium baryon
number violation ceases.)
Once one picks an extension of the standard model which makes
the transition strongly first order, one can compute
$\Delta \sigma^2$.  Here, we will simply take $\Delta \sigma^2 = v^2/2$,
which is approximately what is obtained in the minimal
standard model with a $35~{\rm GeV}$ Higgs mass\cite{KLRS}.
Putting it all together, we find
\begin{equation}
\frac{n_B}{s} \sim f \,(1\times 10^{-9})\, b\, \frac{(5 {\rm TeV})^2}{M^2}\ .
\label{estimate}
\end{equation}
If we take, for example,
$b\sim \alpha_W$ and $M\sim 1~{\rm TeV}$, the bound (\ref{bound})
can be satisfied and (\ref{estimate}) suggests that a
cosmologically relevant BAU may be generated.
We see that if \CP\ violation is introduced
via the operator ${\cal O}$ with a coefficient $b/M^2$ satisfying
(\ref{bound}), and if the bubble walls are thin, 
then the contribution to the baryon asymmetry
of the universe from local electroweak baryogenesis
can be at an interesting level so long as the fudge factor $f$ is
not smaller than about a tenth.  

We have reached the estimate (\ref{estimate}) 
by arguing that $n_B$ must be proportional to $(b/M^2)\Delta \sigma^2$,
arguing by dimensional analysis that it must be proportional to $\xi^{-3}$,
arguing that the time it takes configurations to cross the
barrier in equilibrium in the high temperature phase is not 
relevant, and lumping our remaining ignorance into
$f$.  We now turn to a discussion of what goes into $f$, and
how the treatments of Sections III and IV are related.
There are many contributions to $f$,
since the treatment leading to the 
estimate (\ref{estimate}) 
is greatly over-simplified.  
First, there are the constants $c_i$, which of course differ for the
different configurations in different
cells of volume $\xi^3$,  and must somehow be averaged over.
Second, using the impulse
approximation is not really justified.  In reality, the wall
does not have zero thickness.  More important, even if the wall
{\it is} thin, the time during which it can affect a configuration
of size $\xi$ is at least $\xi$.  Third, as we have already mentioned,
the treatment in terms of the Lagrangian (\ref{Leq}) only has
a chance of capturing the physics near $\tau=0$, and it is not
at all clear what fraction of configurations satisfy this.
Configurations which happen to be farther away from the
crest of the ridge between vacua when the wall hits
do receive a kick from the wall.  However, even if this kick is
large, it may not be in a suitable direction in configuration
space to be effective.  Configurations far from $\tau=0$
can contribute to $n_B$, but their contribution is hard to compute,
because there is no way to reduce the problem to one of one degree
of freedom far from $\tau=0$.
Fourth, even near the crest of the ridge for a given trajectory
the problem does not really reduce to one degree of freedom.
For the configurations of interest, $\sigma$ is a function of
space and time and the operator ${\cal O}$ and the bubble
wall conspire to affect its dynamics.  We have attempted
to describe the effect by treating $\sigma$ as constant
in space and time on either side of the wall and only changing
at the wall.  This is a caricature at best.
Fifth, by now the problem should be sounding more like that of Section III,
and we must face up to the specific difficulties discussed there.
After the wall has passed, the fields are not yet in thermal
equilibrium and their dynamics is complicated.  This may in
fact yield a further contribution to $n_B$.  It may also,
however, negate some of the contribution estimated 
in (\ref{estimate}) because some configurations kicked across
the barrier in one direction by the passage of the wall may at a later time
wander back across the barrier whence they came.
As we discovered in Section III, an estimate of the magnitude
of these sorts of effects is difficult even for the restricted
class of configurations we considered there.  
To sum up, $f$ is almost certainly less than $1$. Hence,
it would be best to 
use (\ref{estimate}) as an upper bound on $n_B/s$, rather
than as an estimate.

\section{Concluding Remarks}

If the electroweak phase transition is strongly first order,
it is possible that the observed baryon asymmetry of the universe
may have been generated by non-equilibrium processes occurring
as bubble walls sweep through the plasma.  In general, there will be
contributions to the asymmetry both from local baryogenesis
and from nonlocal baryogenesis. 
In the last few years, 
much effort has been devoted to computing the contribution
from nonlocal 
effects\cite{CKN2,CKN4,F&S,GS,JPT,CKV,HN}.  
Recently, the work of Moore and
Turok\cite{MT} has made it clear that the computational
resources now available make a large scale numerical treatment
of local baryogenesis a possibility.  Our goal in this paper was
to reconsider two possible routes to a semi-analytic estimate
of the magnitude of the effect.  If the bubble walls are thick,
conditions remain close to thermal equilibrium during the passage
of the wall, and the nonequilibrium physics can be captured by
assigning nonzero chemical potentials to
various quantum numbers including baryon number.  Analytic estimates
for the BAU produced in this setting exist in the 
literature\cite{DHSS,CKN3}
and the need for a numerical treatment is not pressing.
If the bubble walls are thin, however, or if (as is no doubt the case)
they are comparable in thickness to other length scales in the problem,
the situation is unsettled.  We have reanalyzed the problem
of local baryogenesis in the thin wall limit using two different
semi-analytical approaches\cite{TZ,DHSS}.
The method of Dine {\it et al.}
does yield an estimate (\ref{estimate}), but the
uncertainties parametrized by $f$, 
particularly those highlighted by the difficulties
which prevented us from obtaining an estimate via the method
of Turok and Zadrozny, mean that this should be viewed at best
as an upper bound.  A large scale numerical treatment seems necessary.

Moore and Turok\cite{MT} 
have recently taken a big step in this direction.
They have performed $3+1$ dimensional classical 
simulations in which a bubble wall
moves through a box converting the high temperature phase
to the low temperature phase.  To date, they have focused more
on computing quantities like the wall thickness, the wall velocity,
the surface tension, and the drag on the wall and have
only begun their treatment of local electroweak baryogenesis.
To this point, they have introduced \CP\ violation only by
``mocking up'' the effects of (\ref{newop}) by first computing
the average wall profile $\langle \sigma \rangle (z)$ for an ensemble
of walls, and then doing a simulation in which 
one measures the distance of a given point to the nearest bubble
wall and adds 
a chemical potential for Chern-Simons number at that point
proportional to the spatial derivative of the average wall profile
at that distance.  This chemical potential is only nonzero on the
wall, as it would be if it were proportional to $\frac{d}{dt}\sigma^2$
for a moving wall.  
Nevertheless, by imposing the chemical potential
as an external driving force instead of simply introducing (\ref{newop})
in the Lagrangian and letting the dynamics do their thing
self-consistently, one
risks missing a lot of the difficulties (and potential effects)
we have discussed in Section III and at the end of Section IV.  
The simulations of Moore and Turok suggest that 
a large scale numerical assault on the problem of local
electroweak baryogenesis is now possible; the 
difficulties we have discussed which prevent us from
obtaining a reliable semi-analytic estimate
of the magnitude of the effect show that it is necessary. 

Although we have taken considerable care to describe the failings
of (\ref{estimate}), even if we use it only as an upper bound it 
is interesting in the following sense. 
Let us assume that whatever \CP\ violation
is introduced in order to make baryogenesis possible can
be parametrized by the \CP\ violating operator ${\cal O}$ of (\ref{newop}). 
Combined with the experimental
bound (\ref{bound}) on the coefficient of 
${\cal O}$, the result (\ref{estimate}) shows that if
the experimental sensitivity to the electric dipole moment of the
electron or the neutron can be improved by about an order of magnitude,
and if these experiments continue to yield results consistent
with zero, then the baryon asymmetry
of the universe produced by local electroweak baryogenesis 
is smaller than that observed, even if future numerical
simulations were to demonstrate that $f$ is as large as $1$.

\acknowledgments

We would first like to thank Eddie Farhi for fruitful collaboration
during the early stages of this project, and in particular for
his invaluable contributions to Section II.
We would also like to thank the following people for helpful comments and
discussions: Peter Arnold, 
Robert Brandenberger, Sean Carroll, 
Sekhar Chivukula, Jeffrey Goldstone, Ken Johnson, 
Guy Moore, Lisa Randall, Bob Singleton, Neil
Turok, and an anonymous referee.

The work of A.L. and M.T. was supported in part by funds provided by 
the U.S. Department of Energy (D.O.E.) under cooperative research agreement
\# DF-FC02-94ER40818. The work of K.R. was supported in part by the Sherman 
Fairchild Foundation and by the Department of Energy under Grant No.
DE-FG03-92-ER40701.

\end{document}